\documentclass[a4paper,12pt]{article}
\usepackage[utf8]{inputenc}
\usepackage{amsmath}  
\usepackage{amssymb}       
\usepackage{amsfonts}
\usepackage{dsfont}
\usepackage{times}
\usepackage{amsthm} 
\usepackage{epsfig}
\newcommand{\DATUM}{19-Jan-2017}              %                        %
\newcommand{\ol}{\overline}   
  
\newcommand{\Proof}{\noindent\emph{Proof. }}              % Beginning and  %
\newcommand{\QED}{\hspace*{\fill}\mbox{$\Box$}}           % end of a Proof %
\newcommand{\Om}{\Omega}                %%%%%%%%%%%%%%%%%%%%%%%%%%%%%%%%%%%
\newcommand{\om}{\omega}

\newcommand{\rmC}{{\mathrm{C}}}         %%%%%%%%%%%%%%%%%%%%%%
\newcommand{\rmI}{{\mathrm{I}}}         %                    %
\newcommand{\rmR}{{\mathrm{R}}}         %   roman letters    %
\newcommand{\rmS}{{\mathrm{S}}}         %                    %
\newcommand{\one}{\mathbf{1}}

\newcommand{\cB}{\mathcal{B}}
\newcommand{\cD}{\mathcal{D}}

\newcommand{\cF}{\mathcal{F}}

\newcommand{\cH}{\mathcal{H}}

\newcommand{\cL}{\mathcal{L}}         %%%%%%%%%%%%%%%%%%%%%%%%
\newcommand{\cO}{\mathcal{O}}         %                      %
\newcommand{\field}[1]{\mathbb{#1}}
 
\newcommand{\RR}{\field{R}}     %%%%%%%%%%%%%%%%%%%%%%%%%%% 
\newcommand{\NN}{\field{N}}     % Blackboard Bold Letters %
\newcommand{\CC}{\field{C}}     %                         %
\newcommand{\fh}{\mathfrak{h}}  %%%%%%%%%%%%%%%%%%%%%%%%%%%
\newcommand{\hTheta}{\widehat{\Theta}}

\newcommand{\tQ}{\widetilde{Q}}

\newcommand{\fin}{\mathrm{fin}}
\newcommand{\diag}{\mathrm{diag}}

\newcommand{\Ran}{\mathrm{Ran}}              % Miscelleanous Math Symbols %
\newcommand{\cirS}{\mathop{\bigcirc\kern -.73em {\scriptstyle{\rm S}}}}

\newcommand{\hf}{H_\rmR}

\newtheorem{theorem}{Theorem}%[section]        %%%%%%%%%%%%%%%%%%%%%%%%%%%%%%
\newtheorem{lemma}[theorem]{Lemma}             % These Theoremlike environm.%
\newtheorem{corollary}[theorem]{Corollary}     % are counted according to   %
\newtheorem{definition}[theorem]{Definition}   % the section where they     %
\theoremstyle{plain}
\begin{document}
\bibliographystyle{plain}
%%%%%%%%%%%%%%%%%%%%%%%%%%%%%%%%%%%%%%%%%%%%%%%%%%%%%%%%%%%%%%%%%%%%%%%%%%%%
\title{Suppression of Decoherence \\ of a Spin-Boson System \\
  by Time-Periodic Control}

\author{Volker~Bach $<$v.bach@tu-bs.de$>$, \\
Alexander~Hach $<$a.hach@tu-bs.de$>$, \\[1ex]
Institut für Analysis und Algebra \\
TU Braunschweig \\ Pockelsstr.~14 \\
38106 Braunschweig \\ Germany}

\date{\DATUM}

\maketitle

\begin{abstract}
  \noindent We consider a finite-dimensional quantum system coupled to the bosonic radiation field and subject to a time-periodic control operator. Assuming the validity of a certain dynamic decoupling condition we approximate the system's time evolution with respect to the non-interacting dynamics. For sufficiently small coupling constants $g$ and control periods $T$ we show that a certain deviation of coupled and uncoupled propagator may be estimated by $\cO(gt \, T)$. Our approach relies on the concept of Kato stability and general theory on non-autonomous linear evolution equations. 
\end{abstract}


\thispagestyle{empty}

\newpage
\setcounter{page}{1}

%%%%%%%%%%%%%%%%%%%%%%%%%%%%%%%%%%%%%%%%%%%%%%%%%%%%%%%%%%%%%%%%%%%%%%%%%%
%%%%%%%%%%%%%%%%%%%%%%%%%%%%%%%%%%%%%%%%%%%%%%%%%%%%%%%%%%%%%%%%%%%%%%%%%%
%%%%%%%%%%%%%%%%%%%%%%%%%%%%%%%%%%%%%%%%%%%%%%%%%%%%%%%%%%%%%%%%%%%%%%%%%%
\section{Result and Discussion} \label{sec-1}
%%%%%%%%%%%%%%%%%%%%%%%%%%%%%%%%%%%%%%%%%%%%%%%%%%%%%%%%%%%%%%%%%%%%%%%%%%
%%%%%%%%%%%%%%%%%%%%%%%%%%%%%%%%%%%%%%%%%%%%%%%%%%%%%%%%%%%%%%%%%%%%%%%%%%
%%%%%%%%%%%%%%%%%%%%%%%%%%%%%%%%%%%%%%%%%%%%%%%%%%%%%%%%%%%%%%%%%%%%%%%%%%
%
We consider an open quantum system consisting of a small,
finite-dimensional system $\rmS$ coupled to a reservoir $\rmR$ with
infinitely many degrees of freedom. 

Specifically, we assume the small system to be an $N$-level atom, for
some $N \geq 2$, i.e., the system's Hilbert space is $\cH_\rmS =
\CC^N$, with a dynamics generated by a self-adjoint Hamiltonian matrix
\begin{align} \label{eq-1.01}
H_\rmS \ = \ \diag\big( E_{N-1}, E_{N-2}, \ldots, E_1, E_0 \big) ,
\end{align}
which we assume to be diagonal with nonnegative, nondegenerate
eigenvalues $E_{N-1} > E_{N-2} > \ldots > E_1 > E_0 \geq 0$.

The reservoir Hilbert space $\cF_\rmR = \cF_b[\fh]$ is the boson
  Fock space over the square-integrable functions $\fh := L^2(\RR^3)$
  on $\RR^3$ and carries a three-dimensional, massless scalar quantum
  field -- a caricature of the photon field -- whose dynamics is
  generated by the second quantization 
\begin{align} \label{eq-1.02}
\hf \ := \ d\Gamma(\om) \ = \ \int_{\RR^3} \, \om(k) \: a_k^* a_k \: d^3k 
\end{align}
of (the operator of multiplication by) the photon dispersion 
$\om(k) := |k|$. Here, $\{ a_k, a_k^* \}_{k \in \RR^3}$ defines the
standard Fock representation of the canonical commutation relation
(CCR)
\begin{align} \label{eq-1.03}
[ a_p \: , \: a_k] \; = \; [ a_p^* \: , \: a_k^*] \; = \; 0, 
\quad 
[ a_p, a_k^*] \; = \; \delta(p-k), 
\quad 
a_k \Om \; = \; 0,
\end{align}
for all $k, p \in \RR^3$, as an operator-valued distribution, with
$\Om \in \cF_\rmR$ being the normalized vacuum vector.

The Hilbert space of the composite atom-photon system $\rmS + \rmR$ is
the tensor product space $\cH_{\rmS \rmR} = \cH_\rmS \otimes
\cF_\rmR$. Without interaction between these two components, the
dynamics is generated by the self-adjoint Hamiltonian
\begin{align} \label{eq-1.04}
H_{\rmS \rmR}^{(0)} 
\ := \ 
H_\rmS + \hf ,
\end{align}
where here and henceforth we leave out trivial tensor factors whenever
possible and identity, e.g., $H_\rmS \equiv H_\rmS \otimes \one_\rmR$
and $\hf \equiv \one_\rmS \otimes \hf$.

A dipole-type interaction $gH_\rmI$ couples the $N$-level atom to the large
reservoir, i.e., the full, interacting dynamics is generated by the
self-adjoint Hamiltonian
\begin{align} \label{eq-1.05}
H_{\rmS \rmR}^{(g)} \ := \ H_{\rmS \rmR}^{(0)} + g H_\rmI .
\end{align}
Here, $g>0$ is a small coupling constant and
\begin{align} \label{eq-1.06}
H_\rmI \ := \ 
Q \otimes \phi(f) 
\ \equiv \ 
Q \, \phi(f) 
\ = \  
Q \, \big( a^*(f) + a(f) \big)  
\end{align}
is the self-adjoint interaction operator specified by a self-adjoint
complex $N \times N$-matrix $Q = Q^*$ times the field operator
$\phi(f)$. Furthermore, for $f \in \fh$, 
\begin{align} \label{eq-1.07}
a^*(f) \ := \ \int f(k) \; a_k^* \; d^3k, 
\qquad
a(f) \ := \ \int \ol{f(k)} \; a_k \; d^3k .
\end{align}
We assume that $f, \om^{-1} f \in \fh$ which implies the
semiboundedness and self- \linebreak adjointness of $H_{\rmS \rmR}^{(g)}$ on the
domain of $H_{\rmS \rmR}^{(0)}$, for any $g >0$, since under this
assumption $H_\rmI$ is an infinitesimal perturbation of 
$H_{\rmS \rmR}^{(0)}$.

Thanks to the self-adjointness of $H_{\rmS \rmR}^{(g)}$, the evolution
operator it generates according to Schrödinger, i.e., the solution of
the initial value problem 
$\partial_t U_{\rmS \rmR}^{(g)}(t) = -iH_{\rmS \rmR}^{(g)} U_{\rmS \rmR}^{(g)}(t)$, 
$U_{\rmS \rmR}^{(g)}(0) = \one$, is the
strongly continuous one-parameter unitary group $t \mapsto
\exp[-i t H_{\rmS \rmR}^{(g)}]$. Given an initial state of the atom-photon
system by a density matrix $\rho_0 \in \cL_+^1(\cH)$, i.e., a positive
operator of unit trace, the state at time $t \in \RR$ is given by
\begin{align} \label{eq-1.08}
\rho_t \ = \ 
\exp[-i t H_{\rmS \rmR}^{(g)}] \; \rho_0 \; \exp[i t H_{\rmS \rmR}^{(g)}] .
\end{align}
Any initial state $\rho_0$ eventually evolves into the ground state or
the thermal equilibrium state at zero or positive temperature,
respectively, as time $t \to \infty$ grows large. This phenomenon is
usually refered to as \textit{return to equilibrium}. As a
consequence, after a sufficiently long time has elapsed, the state
becomes incoherent and any information initially encoded in it is
lost. A quantum computer can only process data reliably if its
calculations are finished long before the loss of coherence due to the
dissipative process of return to equilibrium described above sets in.

Further perturbations additionally acting on the system would
typically speed up the decoherence process. If the perturbation is
suitably designed, however, the opposite effect might occur and
decoherence is suppressed by the perturbation, rather than enhanced.

The present paper is devoted to the question under which conditions
this suppression of decoherence occurs. More specifically, we study
the influence of a time-periodic perturbation represented by a control
operator $H_\rmC(t)$ which acts on the small system $\rmS$ only. This
latter restriction is a minimal requirement for a physically realistic
model: $H_\rmC(t)$ cannot change the environment. The control operator is
assumed to be a continuous family $H_\rmC \in C[\RR ; \cB(\cH_\rmS)]$ of
self-adjoint complex $N \times N$ matrices such that $H_\rmC(t + T) =
H_\rmC(t)$, for some time period $T>0$ and all $t \in \RR$. 

Acting on the small system as an external force, the generator of the
full dynamics including the control operator $H_\rmC(t)$ is
\begin{align} \label{eq-1.11}
H_{\rmS \rmR \rmC}^{(g)}(t) 
\ := \ 
H_{\rmS \rmR}^{(g)} + H_\rmC(t)
\ = \ 
H_\rmS + \hf + H_\rmC(t) + g H_\rmI .
\end{align}
The theory of non-autonomous linear evolution equations ensures that
for the corresponding time-dependent Schrödinger equation
\begin{align} \label{eq-1.12}
\begin{cases} 
\partial_t U_{\rmS \rmR \rmC}^{(g)}(t,s) 
\ = \ 
-i H_{\rmS \rmR \rmC}^{(g)}(t) \, U_{\rmS \rmR \rmC}^{(g)}(t,s) ,  
& \quad 
U_{\rmS \rmR \rmC}^{(g)}(s,s) \ = \ \one ,
\\[1ex]  
\partial_s U_{\rmS \rmR \rmC}^{(g)}(t,s) 
\ = \ 
i U_{\rmS \rmR \rmC}^{(g)}(t,s) \, H_{\rmS \rmR \rmC}^{(g)}(s) ,  
& \quad 
U_{\rmS \rmR \rmC}^{(g)}(t,t) \ = \ \one ,
\end{cases}
\end{align}
there exists a unique family $U_{\rmS \rmR \rmC}^{(g)} \in C^1[\Delta;
\cB(\cH_{\rmS \rmR})]$ of unitary operators on $\cH_{\rmS \rmR}$,
where $\Delta := \{ (s,t) \in \RR^2 | s \leq t \} \subseteq \RR^2$,
that solves \eqref{eq-1.12}.

Our main result is Theorem~\ref{thm-1.01} below which
asserts that, under Decoupling Condition~\eqref{eq-1.10}, the
deviation of $U_{\rmS \rmR \rmC}^{(g)}(t,0)$ from the identity is
of order $\cO(gt \, T)$, for times smaller than $g^{-1}$. This
is to be compared to the deviation of $\exp[-i t H_{\rmS \rmR}^{(g)}]$
from the identity which is of order $\cO(gt)$. So, for sufficiently
small time periods $T >0$, the control operator effectively slows 
down the evolution and hence also the decoherence of the system.
 
To formulate the decoupling condition, we denote by 
$U_\rmC \in C^1[\Delta; \cB(\cH_\rmS)]$ the propagator generated by
$H_\rmC(t)$, i.e., the unique solution of 
\begin{align} \label{eq-1.09}
\begin{cases} 
\partial_t U_\rmC(t,s) \ = \ -i H_\rmC(t) \, U_\rmC(t,s) ,  & 
\quad U_\rmC(s,s) \ = \ \one ,
\\[1ex] 
\partial_s U_\rmC(t,s) \ = \ i U_\rmC(t,s) \, H_\rmC(s) ,  & 
\quad U_\rmC(t,t) \ = \ \one ,
\end{cases}
\end{align}
and $\tQ(\tau) := U_\rmC(\tau,0) \, Q \, U_\rmC(\tau,0)^*$ on $\cH_\rmS$. 
Our main result is as follows.
%
%%%%%%%%%%%%%%%%%%%%%%%%%%%%%%%%%%%%%%%%%%%%%%%%%%%%%%%%%%%%%%%%%%%%%%%%%%
\begin{theorem} \label{thm-1.01}
Let $L \in \NN$, assume that $\om^{-1} f \in \fh$ and 
$\om^{L+2} f \in \fh$, and set 
\begin{align} \label{eq-1.09,1}
M \ := \ & 2 \, \big\| (\om^{-1/2}+1) f \big\|_2 
\: + \: 2 \, \sum_{k=1}^{L+2} \binom{L+2}{k} \big\| (\om^k+1) f \big\|_2 ,
\\[1ex] \label{eq-1.09,2}
C_{\rmS\rmR\rmC}^{(0)} \ := \ & 
1 \: + \: \| H_\rmS \| \: + \: \sup_{0 \leq r \leq T} \| H_\rmC(r) \| .
\end{align}
Further assume 
%that the time period is $T \leq 1$ %  
that $g \|Q\|  M  T\leq 1$
and that the following 
\textbf{decoupling condition}
\begin{align} \label{eq-1.10}
\int_0^{T} \tQ(\tau) \: d\tau 
\ = \ 
\int_0^{T} U_\rmC(\tau,0) \, Q \, U_\rmC(\tau,0)^* \: d\tau 
\ = \ 0 
\end{align}
holds true. Then, for any $t \geq 0$  and with $n := \left \lfloor \tfrac{t}{T} \right \rfloor, \; \delta := t-n \cdot T$ as well as $\widetilde{C}_g := M \, \|Q\| \: g \cdot \max \lbrace 1, 4 \, C_{\rmS\rmR\rmC}^{(0)} + 3 \, M \, \|Q\| \: g \rbrace$,
%and any $0 \leq g \leq M^{-1} \|Q\|^{-1}$,%

%
\begin{align} \label{eq-1.13}
\Big\| (\hf + \one)^L \,  &
\big( U_{\rmS \rmR \rmC}^{(g)}(t,0)  - U_{\rmS \rmR \rmC}^{(0)}(t,0) \big) \, 
(\hf + \one)^{-L-2} \Big\|
\nonumber \\[1ex]
& \ \leq   \
T \cdot M \, \|Q\| \: g \left[ \delta + \left(4 \, C_{\rmS\rmR\rmC}^{(0)} + 3 \, M \, \|Q\| \: g \right) n T \right] \, \exp[\|Q\| \, M \, gt ] 
\nonumber \\[1ex] 
& \ \leq   \
T \cdot \widetilde{C}_g \, t \, \exp[\|Q\| \, M \, gt ] .
\end{align}
\end{theorem}
%%%%%%%%%%%%%%%%%%%%%%%%%%%%%%%%%%%%%%%%%%%%%%%%%%%%%%%%%%%%%%%%%%%%%%%%%%
%
We discuss Theorem~\ref{thm-1.01}:
\begin{itemize}
\item The idea of suppression of decoherence by a periodic control
  goes back to \cite{Facchi}. Theorem~\ref{thm-1.01} was proven with
  mathematical rigor in ~\cite{BachPedraMerkliSigal2014}, but under
  stronger assumptions and with considerably more involved methods:
  \begin{itemize}
  \item[-] First, the reservoir in \cite{BachPedraMerkliSigal2014} was
    assumed to represent a fermion, rather than a boson field.

  \item[-] Secondly, the control operator $H_\rmC(t)$ was assumed to
    commute with the Hamiltonian $H_\rmS$ of the atom, $[H_\rmC(t),
    H_\rmS] = 0$, for all $t \in \RR$.

  \item[-] A third difference is the framework of Liouvilleans as
    generators of the dynamics at nonzero temperatures which is
    considerably more involved on a technical level. 

  \item[-] On the other hand, the approach in
    \cite{BachPedraMerkliSigal2014} yields control on the dynamics for
    all times -- large and small -- and, in particular, allows to
    follow the rate of convergence to the limiting state, as $t \to
    \infty$. In contrast, the methods used in the present paper give
    nontrivial estimates only for times less than $g^{-1}$, which is
    large compared to unity but small compared to the van Hove time
    scale $\sim g^{-2}$.
  \end{itemize}

\item We observe that Decoupling Condition~\eqref{eq-1.10} and
  $\partial_t \tQ(t) = - i [ H_\rmC(t) , \tQ(t) ]$ imply
\begin{align} \label{eq-1.14}
-T \, \tQ(0) 
\ = \ 
\int_0^T dt \, \big( \tQ(t) - \tQ(0) \big)
\ = \ 
- i \int_0^T dt \int_0^t ds \, [ H_\rmC(s) , \tQ(s) ] .
\end{align}
Since $\| \tQ(s) \| = \|Q\|$, for all $s \in [0,T]$, the triangle
inequality hence yields
\begin{align} \label{eq-1.15}
\int_0^T \| H_\rmC(t) \| \: dt \ \geq \ \frac{1}{2} .
\end{align}
This estimate shows that due to Decoupling Condition~\eqref{eq-1.10}
the action the control operator excerts on the system in a single cycle
is at least of the order of unity with respect to natural units ($\hslash=1$). Assuming a control period $T$ corresponding to a physically feasible time resolution of a hypothetical control operator $H_\rmC(t)$, e.g. a femtosecond regime $T \sim 10^{-15}\, \mathrm{s}$, the energy density in SI-units
of such a device acting on an atom-sized quantum system would be about $10^{11}\, \mathrm{J}/\mathrm{m}^3$.

%\item Moreover, to yield a nontrivial suppression of decoherence
%according to \eqref{eq-1.13} the time period has to be small compared
%to unity, $T \ll 1$, which in SI-units would be \ref{time}~Seconds.  
\end{itemize}

In the following Section~\ref{sec-2} we review some standard material
on solutions of linear non-autonomous evolution equations on Banach
spaces for which we focus on the special case of unitary propagators
for the time-dependent Schrödinger equation. In order to apply this
theory to the present model situation of a spin-boson model with a
time-periodic control, we then derive the necessary relative operator
bounds. After these preparations, we proceed to the proof of
Theorem~\ref{thm-1.01} given in Section~\ref{sec-3}.

\newpage

%%%%%%%%%%%%%%%%%%%%%%%%%%%%%%%%%%%%%%%%%%%%%%%%%%%%%%%%%%%%%%%%%%%%%%%%%%
%%%%%%%%%%%%%%%%%%%%%%%%%%%%%%%%%%%%%%%%%%%%%%%%%%%%%%%%%%%%%%%%%%%%%%%%%%
%%%%%%%%%%%%%%%%%%%%%%%%%%%%%%%%%%%%%%%%%%%%%%%%%%%%%%%%%%%%%%%%%%%%%%%%%%
\section{Propagators and Kato Stability} \label{sec-2}
%%%%%%%%%%%%%%%%%%%%%%%%%%%%%%%%%%%%%%%%%%%%%%%%%%%%%%%%%%%%%%%%%%%%%%%%%%
%%%%%%%%%%%%%%%%%%%%%%%%%%%%%%%%%%%%%%%%%%%%%%%%%%%%%%%%%%%%%%%%%%%%%%%%%%
%%%%%%%%%%%%%%%%%%%%%%%%%%%%%%%%%%%%%%%%%%%%%%%%%%%%%%%%%%%%%%%%%%%%%%%%%%
%
In this section we recall a standard set of sufficient conditions for
the existence of a (unitary) propagator 
$\big( U(t,s) \big)_{(t,s) \in \Delta}$ for the time-dependent
Schrödinger equation
\begin{align} \label{eq-2.01}
\forall \, (t,s) \in \Delta : \quad 
\begin{cases} 
\partial_t U(t,s) \; = \; -i H(t) \, U(t,s),  
& \quad U(s,s) \; = \; \one ,
\\[1ex] 
\partial_s U(t,s) \; = \; iU(t,s) \, H(t), 
& \quad U(t,t) \; = \; \one ,
\end{cases}
\end{align}
given by using the concept of \textit{Kato quasi-stability}.

To define this notion we assume $\big(X, \| \cdot \| \big)$ to be a
complex Banach space with a dense Banach subspace $Y \subseteq X$
whose norm $\| \cdot \|_Y$ can be written as 
$\| x \|_Y = \| \hTheta x \|$ for a suitable linear, isometric
bijection $\hTheta: Y \to X$. We further assume that 
$\| \hTheta x \| \geq \|x\|$, for all $x \in X$. The operator $\hTheta$
allows us to avoid using the norm $\| \cdot \|_Y$ altogether.
%
%%%%%%%%%%%%%%%%%%%%%%%%%%%%%%%%%%%%%%%%%%%%%%%%%%%%%%%%%%%%%%%%%%%%%%%%%%
\begin{definition} \label{def-2.01}
Let $\big(X, | \cdot | \big)$ be a complex Banach space and 
$Y \subseteq X$ a dense Banach subspace. A family 
$G \equiv \big( G(t) \big)_{t \in \RR_0^+}$ 
of densely defined, closed operators $G(t)$ is called 
\textbf{Kato quasi-stable}, if there exists a constant 
$C \geq 1$ and continuous maps $\beta_0, \beta_1: \RR_0^+ \to \RR_0^+$ 
such that following conditions~B1, B2, and B3 are satisfied:
\begin{itemize}
\item[B1] The operators $G$ define a norm-continuous family of bounded 
operators from $Y$ to $X$, i.e., 
$G \hTheta^{-1} \in C\big[\RR_0^+,\cB(X)\big]$.

\item[B2] The commutators 
$[\hTheta, G(t)] \hTheta^{-1} := \hTheta G(t) \hTheta^{-1} - G(t)$ are 
densely defined on $X$ and extend to a continuous family of bounded
operators, $[\hTheta, G(t)] \hTheta^{-1} \in C\big[\RR_0^+,\cB(X)\big]$, 
with $\big\| [\hTheta, G(t)] \hTheta^{-1} \big\|_{\cB(X)} = \beta_1(t)$.

\item[B3] For all $n\in\NN$, all $t_1, \ldots, t_n \in \RR_0^+$, and all 
$\lambda_1 > \beta_0(t_1), \ldots, \lambda_n > \beta_0(t_n)$, the norm 
estimate
\begin{align} \label{eq-2.02}
\bigg| \prod_{k=1}^n \big( \lambda_k - G(t_k) \big)^{-1} \bigg|
\ \leq \ 
C \cdot \prod_{k=1}^n \frac{1}{\lambda_k - \beta_0(t_k)}
\end{align}
holds true.
\end{itemize}
\end{definition}
%%%%%%%%%%%%%%%%%%%%%%%%%%%%%%%%%%%%%%%%%%%%%%%%%%%%%%%%%%%%%%%%%%%%%%%%%%
%
One of the main results of the theory on non-autonomous linear
evolution equations is Theorem~\ref{thm-2.02}, below; see, e.g.,
\cite{EngelNagel,BachBru2010,BachBru2016}. A key element in the proof of
Theorem~\ref{thm-2.02} in \cite{BachBru2010} and in \cite{BachBru2016} is
the \textit{Yosida approximation} $G_\lambda(t) := -\lambda +
\lambda^2 [\lambda - G(t)]^{-1}$, for $\lambda > \beta_0(t)$, which
defines a family of bounded operators that strongly converge to
$G(t)$, as $\lambda \to \infty$.
%
%%%%%%%%%%%%%%%%%%%%%%%%%%%%%%%%%%%%%%%%%%%%%%%%%%%%%%%%%%%%%%%%%%%%%%%%%%
\begin{theorem} \label{thm-2.02}
Let $\big(X, | \cdot | \big)$ be a complex Banach space, 
$Y \subseteq X$ a dense Banach subspace, and
$G \equiv \big( G(t) \big)_{t \in \RR_0^+}$ a Kato quasi-stable family
of densely defined, closed operators, with 
$M \geq 1$, $\beta_0, \beta_1: \RR_0^+ \to \RR_0^+$ 
corresponding to Conditions~B1, B2, and B3.
Then there exists a unique solution
$\big( U(t,s) \big)_{(t,s) \in \Delta}$ for the non-autonomous linear
evolution equation
\begin{align} \label{eq-2.03}
\forall \, (t,s) \in \Delta : \quad 
\begin{cases} 
\partial_t U(t,s) \; = \; G(t) \, U(t,s),  
& \quad U(s,s) \; = \; \one,
\\[1ex] 
\partial_s U(t,s) \; = \; - U(t,s) \, G(t), 
& \quad U(t,t) \; = \; \one,
\end{cases}
\end{align}
which obeys the following norm bounds,
\begin{align} \label{eq-2.04}
\| U(t,s) \|
\ \leq \ &
C \: \int_s^t \beta_0(\tau) \, d\tau, 
\\[1ex] \label{eq-2.05}
\big\| \hTheta \, U(t,s) \, \hTheta^{-1} \big\|
\ \leq \ &
C \: \int_s^t \big\{ \beta_0(\tau) + C \, \beta_1(\tau) \big\} \, d\tau, 
\end{align}
for all $(t,s) \in \Delta$.
\end{theorem}
%%%%%%%%%%%%%%%%%%%%%%%%%%%%%%%%%%%%%%%%%%%%%%%%%%%%%%%%%%%%%%%%%%%%%%%%%%
%
If $X$ is specified to be a complex Hilbert space $\cH$, 
$\cD = \Ran(\hTheta) \subseteq \cH$, for some unbounded, self-adjoint 
operator $\hTheta \geq \one$, and $G$ is a strongly continuous family 
$-iH \equiv \big( -iH(t) \big)_{t \in \RR_0^+}$ of skew-adjoint
operators on $\cH$, then Condition~B3 in Definition~\ref{def-2.01} is
automatic with $C = 1$ and $\beta_0 \equiv 0$, and 
Theorem~\ref{thm-2.02} can be strengthened to the following assertion.
%
%%%%%%%%%%%%%%%%%%%%%%%%%%%%%%%%%%%%%%%%%%%%%%%%%%%%%%%%%%%%%%%%%%%%%%%%%%
\begin{theorem} \label{thm-2.03}
Let $\cH$ be a separable complex Hilbert space, 
$\cD = \Ran(\hTheta) \subseteq \cH$, for some unbounded, 
self-adjoint  operator $\hTheta \geq \one$, and 
$H \equiv \big( H(t) \big)_{t \in \RR_0^+}$ a strongly continuous
family of self-adjoint operators $H(t) = H^*(t)$ on $\cH$ such that 
$H \hTheta^{-1} , [\hTheta, H(t)] \hTheta^{-1} \in C\big[\RR_0^+,\cB(\cH)\big]$.
Then there exists a unique propagator 
$\big( U(t,s) \big)_{(t,s) \in \Delta}$ to the time-dependent
Schrödinger equation
\begin{align} \label{eq-2.06}
\forall \, (t,s) \in \Delta : \quad 
\begin{cases} 
\partial_t U(t,s) \; = \; -i H(t) \, U(t,s),  
& \quad U(s,s) \; = \; \one ,
\\[1ex] 
\partial_s U(t,s) \; = \; iU(t,s) \, H(t), 
& \quad U(t,t) \; = \; \one ,
\end{cases}
\end{align}
which is a family of unitary operators fulfilling
the norm estimate
\begin{align} \label{eq-2.07}
\big\| \hTheta \, U(t,s) \, \hTheta^{-1} \big\|
\ \leq \ &
\int_s^t \big\| [\hTheta, G(\tau)] \hTheta^{-1} \big\| \, d\tau, 
\end{align}
for all $(t,s) \in \Delta$.
\end{theorem}
%%%%%%%%%%%%%%%%%%%%%%%%%%%%%%%%%%%%%%%%%%%%%%%%%%%%%%%%%%%%%%%%%%%%%%%%%%
%
To apply Theorem~\ref{thm-2.03} to the present model situation,
we choose 
\begin{align} \label{eq-2.08}
\cH \; := \; \cH_{\rmS\rmR} , \ \
\hTheta \; := \; \Theta^{L+2} , \ \ 
\Theta \; := \; \hf + \one , \ \ 
H(t) \; := \; H_{\rmS\rmR\rmC}(t) .
\end{align}
To validate the hypothesis of Theorem~\ref{thm-2.03}, we define
\begin{align} \label{eq-2.09a}
M_{-1/2} \ := \ &
2 \: \big\| (\om^{-1/2}+1) f \big\|_2 ,
\\[1ex] \label{eq-2.09b}
M_n \ := \ &
2 \: \sum_{k=1}^n \binom{n}{k} \big\| (\om^k+1) f \big\|_2 
\end{align}
and establish the following bounds.
%
%%%%%%%%%%%%%%%%%%%%%%%%%%%%%%%%%%%%%%%%%%%%%%%%%%%%%%%%%%%%%%%%%%%%%%%%%%
\begin{lemma} \label{lem-2.04}
Let $n \in \NN$ and assume that $\om^{-1/2} f, \om^n f \in L^2(\RR^3)$.
Then 
\begin{align} \label{eq-2.10}
\| a(f) \Theta^{-1} \| , \ \| a^*(f) \Theta^{-1} \| 
\ \leq \ &
\frac{1}{2} \, M_{-1/2}.
\\[1ex] \label{eq-2.11}
\big\| [\Theta^n , a^*(f)] \Theta^{-n} \big\|, \
\big\| [\Theta^n , a(f)] \Theta^{-n} \big\|
\ \leq \ &
\frac{1}{2} \, M_n .
\end{align}
\end{lemma}
%%%%%%%%%%%%%%%%%%%%%%%%%%%%%%%%%%%%%%%%%%%%%%%%%%%%%%%%%%%%%%%%%%%%%%%%%%
%
\Proof 
It is convenient to introduce the subspace 
$\cF_\rmR^\fin \subseteq \cF_\rmR$ of \textit{finite vectors} whose
elements have only finitely many non-vanishing components, each being
smooth and compactly supported. For any normalized finite vector 
$\psi \in \cF_\rmR^\fin$, we have that
\begin{align} \label{eq-2.12}
\| a^*(\om^k f) \psi\|^2
\ = \ 
\| \om^k f \|_2^2 \, + \, \| a(\om^k f) \psi\|^2  , 
\end{align}
for all $k \geq 0$. Additionally requiring that $k \geq 1$, we further
have
\begin{align} \label{eq-2.13}
\| a(\om^k f) \psi\|
\ \leq \ & 
\int |f(\xi)| \: \| \om^k(\xi) a_\xi \psi \| \: d\xi
\ \leq \ 
\|f\|_2 \cdot 
\big\langle \psi \big| d\Gamma[\om^{2k}] \, \psi \big\rangle^{1/2}
\\[1ex] \nonumber 
\ \leq \ &
\|f\|_2 \cdot 
\big\langle \psi \big| \big( d\Gamma[\om] \big)^{2k} 
\, \psi \big\rangle^{1/2}
\ = \ 
\|f\|_2 \cdot \| \hf^k \psi \| 
\ \leq \ 
\|f\|_2 \cdot \| \Theta^k \psi \| ,
\end{align}
where $d\Gamma(A)$ denotes the second quantization of an operator
$A$. For $k=0$, we slightly modify this estimate and obtain
\begin{align} \label{eq-2.14}
\| a(f) \psi\|
\ \leq \ 
\int |f(\xi)| \: \| a_\xi \psi \| \: d\xi
\ \leq \ 
\big\| \om^{-1/2} f \big\|_2 \cdot \| \hf^{1/2} \psi \| .
\end{align}
This estimate and \eqref{eq-2.12} with $k=0$ establish
\begin{align} \label{eq-2.15}
\| a(f) \Theta^{-1} \| , \ \| a^*(f) \Theta^{-1} \| 
\ \leq \ 
\big\| (\om^{-1/2} + 1) f \big\|_2 
\end{align}
and hence \eqref{eq-2.10}.

On the other hand, Eq.~\eqref{eq-2.12} and \eqref{eq-2.13} imply
for $k \geq 1$ that
\begin{align} \label{eq-2.16}
\| a(\om^k f) \, \Theta^{-k} \|, \ \| a^*(\om^k f) \, \Theta^{-k} \|
\ \leq \  
\big\| (\om^k + 1) f \big\|_2 .
\end{align}
Using the identities
\begin{align} \label{eq-2.17}
\big[ \hf , a^*(f) \big] \ = \ a^*(\om f), \quad
\big[ \hf , a(f) \big] \ = \ -a(\om f), 
\end{align}
and an induction, we easily find that
\begin{align} \label{eq-2.18}
\Theta^n \, a^*(f) \, \Theta^{-n}
\ = \ &
\Theta^{n-1} \, a^*(f) \, \Theta^{-(n-1)}
\: + \: 
\Theta^{n-1} \, a^*(\om f) \, \Theta^{-n}
\ = \ \ldots
\nonumber \\[1ex]
\ = \ & 
\sum_{k=0}^n \binom{n}{k} \: a^*(\om^k f) \, \Theta^{-k} ,
\end{align}
and similarly 
\begin{align} \label{eq-2.19}
\Theta^n \, a(f) \, \Theta^{-n}
\ = \ 
\sum_{k=0}^n (-1)^k \: \binom{n}{k} \: a(\om^k f) \, \Theta^{-k} .
\end{align}
Putting \eqref{eq-2.18}, \eqref{eq-2.19} and \eqref{eq-2.16} together,
we obtain
\begin{align} \label{eq-2.20}
\big\| [\Theta^n , a^*(f)] \Theta^{-n} \big\|, \
\big\| [\Theta^n , a(f)] \Theta^{-n} \big\|
\ \leq \ 
\sum_{k=1}^n \binom{n}{k} \: \big\| (\om^k +1) \, f \big\|_2 .
\end{align}
\QED

Since $\tau \mapsto H_{\rmC}(\tau)$ is continuous and 
$[ H_{\rmS\rmR\rmC}(\tau), \Theta] = g Q \otimes [a^*(f) + a(f), \hf]$, 
Lemma~\ref{lem-2.04} and Theorem~\ref{thm-2.03} imply the following
corollary.
%
%%%%%%%%%%%%%%%%%%%%%%%%%%%%%%%%%%%%%%%%%%%%%%%%%%%%%%%%%%%%%%%%%%%%%%%%%%
\begin{corollary} \label{cor-2.05}
Let $L \in \NN_0$ and assume that $\om^{-1/2} f, \om^{L+2} f \in L^2(\RR^3)$.
Then $\tau \mapsto H_{\rmS\rmR\rmC}^{(g)}(\tau) \, \Theta^{-1}$ is continuous 
and bounded, uniformly in $\tau \in \RR$, and fulfills the following 
estimates,
\begin{align} \label{eq-2.21}
\big\| \Theta^\ell \, H_{\rmI} \, \Theta^{-(\ell+j)} \big\| 
\ \leq \ 
\|Q\| \, (M_{-1/2} + M_{\ell+j}) ,
\\[1ex]
 \label{eq-2.22}
\big\| \, [\Theta^{\ell+j} , H_{\rmS\rmR\rmC}^{(g)}(\tau)] \Theta^{-(\ell+j)} \big\| 
\ \leq \ 
g \, \|Q\| \, M_{\ell+j} ,
\end{align}
for all $\tau \in \RR$, all $\ell \in \{0, 1, \ldots, L\}$, and $j \in
\{1, 2\}$. Moreover, $\tau \mapsto H_{\rmS\rmR\rmC}^{(g)}(\tau)$ is a
Kato-quasistable family of self-adjoint operators, and the unique,
unitary solution of
\begin{align} \label{eq-2.23}
\begin{cases} 
\partial_t U_{\rmS \rmR \rmC}^{(g)}(t,s) 
\ = \ 
-i H_{\rmS \rmR \rmC}^{(g)}(t) \, U_{\rmS \rmR \rmC}^{(g)}(t,s) ,  
& \quad 
U_{\rmS \rmR \rmC}^{(g)}(s,s) \ = \ \one ,
\\[1ex]  
\partial_s U_{\rmS \rmR \rmC}^{(g)}(t,s) 
\ = \ 
i U_{\rmS \rmR \rmC}^{(g)}(t,s) \, H_{\rmS \rmR \rmC}^{(g)}(s) ,  
& \quad 
U_{\rmS \rmR \rmC}^{(g)}(s,s) \ = \ \one ,
\end{cases}
\end{align}
obeys
\begin{align} \label{eq-2.24}
\big\| \Theta^{\ell+j} \, U_{\rmS\rmR\rmC}^{(g)}(t,s) \, \Theta^{-(\ell+j)} \big\| 
\ \leq \ 
\exp\big[ g \, \|Q\| \, M_{\ell+j} \, (t-s) \big],
\end{align}
for all $(t,s) \in \Delta$, all $\ell \in \{0, 1, \ldots, L\}$, and $j \in
\{1, 2\}$. .
\end{corollary}
%%%%%%%%%%%%%%%%%%%%%%%%%%%%%%%%%%%%%%%%%%%%%%%%%%%%%%%%%%%%%%%%%%%%%%%%%%
%

\newpage

%%%%%%%%%%%%%%%%%%%%%%%%%%%%%%%%%%%%%%%%%%%%%%%%%%%%%%%%%%%%%%%%%%%%%%%%%%
%%%%%%%%%%%%%%%%%%%%%%%%%%%%%%%%%%%%%%%%%%%%%%%%%%%%%%%%%%%%%%%%%%%%%%%%%%
%%%%%%%%%%%%%%%%%%%%%%%%%%%%%%%%%%%%%%%%%%%%%%%%%%%%%%%%%%%%%%%%%%%%%%%%%%
\section{Proof of Theorem~\ref{thm-1.01}} \label{sec-3}
%%%%%%%%%%%%%%%%%%%%%%%%%%%%%%%%%%%%%%%%%%%%%%%%%%%%%%%%%%%%%%%%%%%%%%%%%%
%%%%%%%%%%%%%%%%%%%%%%%%%%%%%%%%%%%%%%%%%%%%%%%%%%%%%%%%%%%%%%%%%%%%%%%%%%
%%%%%%%%%%%%%%%%%%%%%%%%%%%%%%%%%%%%%%%%%%%%%%%%%%%%%%%%%%%%%%%%%%%%%%%%%%
%
We first fix $n \in \NN$ and $\delta \in [0,1)$ so that $t = nT +
\delta$. We abbreviate
\begin{align} \label{eq-1.16}
\Theta \ := \ \hf + \one , \ \ 
U_\kappa(s) \; := \; U_{\rmS \rmR \rmC}^{(\kappa)}(s,0) , \ \
W(s) \; := \; U_0(s)^* \, U_g(s)  - \one , 
\end{align}
for $\kappa \geq 0$. Next we claim that, for all $s \geq 0$,
\begin{align} \label{eq-1.17}
U_0(s + T) \; = \; U_0(s) \, U_0(T) , \quad
U_g(s + T) \; = \; U_g(s) \, U_g(T) . 
\end{align}
Indeed, $\partial_s U_\kappa(s) \, U_\kappa(T) = 
-i H_{\rmS \rmR \rmC}^{(\kappa)}(s) \, U_\kappa(s) \, U_\kappa(T)$ and
\begin{align} \label{eq-1.18}
\partial_s U_\kappa(s+T) 
\; = \; 
-i H_{\rmS \rmR \rmC}^{(\kappa)}(s+T) \, U_\kappa(s+T) 
\; = \; 
-i H_{\rmS \rmR \rmC}^{(\kappa)}(s) \, U_\kappa(s+T) ,
\end{align}
since $H_{\rmS \rmR \rmC}^{(\kappa)}$ is $T$-periodic. The uniqueness
of the solution of linear ODE with the same initial value then implies
\eqref{eq-1.17} both for $\kappa = 0$ and for $\kappa = g$. 

Eq.~\eqref{eq-1.17} in turn implies that
\begin{align} \label{eq-1.19}
U_g(t) - U_0(t) 
\ = \ & 
U_g(\delta + nT) - U_0(\delta + nT)
\nonumber \\[1ex] 
\ = \ & 
\big[ U_g(\delta) - U_0(\delta) \big] \, U_g(nT) 
\; + \; 
U_0(\delta) \, \big[ U_g(nT) - U_0(nT) \big]  
\nonumber \\[1ex] 
\ = \ & 
U_0(\delta) \, W(\delta) \, U_g(nT)
\\ \nonumber 
& \; + \; 
\sum_{j=1}^n U_0\big( \delta + (j-1)T \big) \, 
\big[ U_g(T) - U_0(T) \big] U_g\big( (n-j) T \big)
\\ \nonumber 
\ = \ & 
U_0(\delta) \, W(\delta) \, U_g(nT)
\; + \; 
\sum_{k=0}^{n-1} U_0\big( \delta + (n-k) T \big) \, W(T) \, U_g(kT) .
\end{align}
Since $U_0(s)$ is unitary and commutes with $\Theta$ this identity
implies that
\begin{align} \label{eq-1.20}
\big\| \Theta^L \, \big( U_g(t) - & U_0(t) \big) \, 
\Theta^{-L-2} \big\|
\nonumber \\[1ex] 
\ \leq \ &
\big\| \Theta^L \, W(\delta) \, U_g(nT) \, \Theta^{-L-2} \big\|
\: + \:
\sum_{k=0}^{n-1} \big\| \Theta^L \, W(T) \, U_g(kT) \, \Theta^{-L-2} \big\|
\nonumber \\[1ex] 
\ \leq \ &
\Big( \big\| \Theta^L \, W(\delta) \, \Theta^{-L-2} \big\| \: + \:
n \cdot \big\| \Theta^L \, W(T) \, \Theta^{-L-2} \big\| \Big)
\nonumber \\ 
& \hspace{10mm} \cdot
\sup_{0 \leq \tau \leq nT} \big\| \Theta^{L+2} \, U_g(\tau) \, \Theta^{-L-2} \big\| .
\end{align}
Thanks to Corollary~\ref{cor-2.05}, Eq.~\eqref{eq-2.24} above, we have that
\begin{align} \label{eq-1.21}
\big\| \Theta^{L+2} \, U_g(\tau) \, \Theta^{-L-2} \big\| 
\ \leq \ &
\exp[ \|Q\| \, M \, g \tau ] ,
\end{align}
for all $\tau \geq 0$, using that $M_{L+2} \leq M$. Furthermore, for
all $0 \leq s \leq T$, the fundamental theorem of calculus gives
\begin{align} \label{eq-1.22}
\Theta^{L+1} \, W(s) \, \Theta^{-L-2}
\ = \ &
i \int_0^s \Theta^{L+1} \, U_0(s)^* \, 
\big( H_{\rmS \rmR \rmC}^{(0)}(t) - H_{\rmS \rmR \rmC}^{(g)}(t) \big) \, 
U_g(s) \, \Theta^{-L-2} \; ds
\nonumber \\[1ex]
\ = \ &
i \, g \, \int_0^s \Theta^{L+1} \, U_0(s)^* \, H_\rmI \, 
U_g(s) \, \Theta^{-L-2} \; ds ,
\end{align}
and with \eqref{eq-1.21} and Corollary~\ref{cor-2.05},
Eq.~\eqref{eq-2.21} this implies that
\begin{align} \label{eq-1.23}
\big\| \Theta^L \, W(\delta) \, \Theta^{-L-2} \big\|
\ \leq \ &
\big\| \Theta^{L+1} \, W(\delta) \, \Theta^{-L-2} \big\|
\nonumber \\[1ex]
\ \leq \ &
g \, \delta \, \big\| \Theta^{L+1} \, H_\rmI \, \Theta^{-L-2} \big\|
\cdot \exp[ \|Q\| \, M \, g \, \delta]
\nonumber \\[1ex]
\ \leq \ &
g \, \delta \, \|Q\| \, M \, \exp[ \|Q\| \, M \, g \, \delta] ,
\end{align}
additionally using that $M = M_{-1/2} + M_{L+2}$.
Inserting \eqref{eq-1.21} and \eqref{eq-1.23} into \eqref{eq-1.20}, 
we obtain
\begin{align} \label{eq-1.24}
\big\| \Theta^L & \, \big( U_g(t) - U_0(t) \big) \, 
\Theta^{-L-2} \big\|
\\[1ex] \nonumber 
& \ \leq \ 
\Big( g \, \delta \, \|Q\| \, M \: + \:
n \cdot \big\| \Theta^L \, W(T) \, \Theta^{-L-2} \big\| \Big)
\cdot 
\exp[ \|Q\| \, M \, g \, t] .
\end{align}
For the estimate of $\| \Theta^L \, W(T) \, \Theta^{-L-2} \|$, we observe
that, again by the fundamental theorem of calculus, 
\begin{align} \label{eq-1.25}
\Theta^L \, W(T) \, \Theta^{-L-2}
\ = \ &
i \, g \, \int_0^T \Theta^L \, U_0(s)^* \, H_\rmI \, 
U_g(s) \, \Theta^{-L-2} \; ds 
\nonumber \\[1ex]
\ = \ &
i \, g \, \int_0^T 
\Theta^L \, U_0(s)^* \, H_\rmI \, U_0(s) \, W(s) \, \Theta^{-L-2} \; ds 
\nonumber \\[1ex]
& \; + \; 
i \, g \, \int_0^T 
\Theta^L \, U_0(s)^* \, H_\rmI \, U_0(s) \, \Theta^{-L-2} \; ds ,
\end{align}
which implies that
\begin{align} \label{eq-1.26}
\big\| \Theta^L \, W(T) \, \Theta^{-L-2} \big\|
\ \leq \ &
g \, \big\| \Theta^L \, H_\rmI \, \Theta^{-L-1} \big\| \;
\int_0^T \big\| \Theta^{L+1} \, W(s) \, \Theta^{-L-2} \big\| \; ds 
\nonumber \\
& \; + \; 
g \, \bigg\| \int_0^T 
\Theta^L \, U_0(s)^* \, H_\rmI \, U_0(s) \, \Theta^{-L-2} \; ds \bigg\|
\nonumber \\[1ex]
\ \leq \ &
g^2 \, T^2 \, \|Q\|^2 \, M^2 \, \exp[ \|Q\| \, M \, g \, T] 
\nonumber \\
& \; + \; 
g \, \bigg\| \int_0^T 
\Theta^L \, U_0(s)^* \, H_\rmI \, U_0(s) \, \Theta^{-L-2} \; ds \bigg\| ,
\end{align}
using \eqref{eq-1.23} and again \eqref{eq-2.21} and 
$M = M_{-1/2} + M_{L+2}$. 

We proceed to the key estimate of this paper whose proof uses
Decoupling Condition~\eqref{eq-1.10}. Namely, we observe that
\begin{align} \label{eq-1.27}
& \int_0^T  
U_0(s)^* \, H_\rmI \, U_0(s) \; ds
\ = \ 
\int_0^T  ds \:
\big\{ U_0(s)^* \, H_\rmI \, U_0(s) 
\: - \:  U_{\rmC}(s)^* \, H_\rmI \, U_{\rmC}(s) \big\}
\nonumber \\[1ex]
& \ = \ 
\int_0^T ds \int_0^s dr \:
\big\{ U_0(r)^* \, 
\big[ H_{\rmS\rmR\rmC}^{(0)}(r) , H_\rmI \big] \, U_0(r) 
\: - \:  
U_{\rmC}(r)^* \, \big[ H_{\rmC}(r) , H_\rmI \big] \, U_{\rmC}(r) \big\} .
\end{align}
Since $U_0(r)$, $U_{\rmC}(r)$, $H_{\rmS}$, $H_{\rmR}$, and
$H_{\rmC}(r)$ all commute with $\Theta$, this, \eqref{eq-2.21}, and
$M = M_{-1/2} + M_{L+2}$
imply that
\begin{align} \label{eq-1.28}
 \bigg\| \int_0^T 
\Theta^L \, & U_0(s)^* \, H_\rmI \, U_0(s) \, \Theta^{-L-2} \; ds \bigg\|
\\[1ex] \nonumber 
\ \leq \ & 
2 \, T^2 \, \Big(
\big\| \Theta^L \, H_\rmI \, \Theta^{-L-1} \big\| 
+ \big\| \Theta^{L+1} \, H_\rmI \, \Theta^{-L-2} \big\| \Big) 
\\ \nonumber 
\ \ & 
\hspace{10mm} \cdot \Big( \big\| H_{\rmR} \Theta^{-1} \big\|
+ \| H_\rmS \| + \sup_{0 \leq r \leq T} \| H_\rmC(r) \| \Big) 
\\[1ex] \nonumber 
\ \leq \ & 
4 \, C_{\rmS\rmR\rmC}^{(0)} \, T^2 \, \|Q\| \, M ,
\end{align}
where $C_{\rmS\rmR\rmC}^{(0)} := 
1 + \| H_\rmS \| + \sup_{0 \leq r \leq T} \| H_\rmC(r) \|$.
Inserting \eqref{eq-1.28} into \eqref{eq-1.26} and the resulting
estimate into \eqref{eq-1.24}, we arrive at the assertion, taking
into account that $g \|Q\|  M  T\leq 1$ 
%and $T \leq 1$
 which implies that $\exp[ \|Q\| \, M \, g \, T] \leq e \leq 3$.
\qed

%%%%%%%%%%%%%%%%%%%%%%%%%%%%%%%%%%%%%%%%%%%%%%%%%%%%%%%%%%%%%%%%%%%%%%%%%%

\inputencoding{latin2}
\bibliographystyle{plain}
\bibliography{MyBib}

\begin{thebibliography}{1}

\bibitem{BachBru2010}
V.~Bach and J.-B. Bru.
\newblock Rigorous foundations of the brockett-wegner flow for operators.
\newblock {\em J. Evol. Equ.}, 10:425--442, 2010.

\bibitem{BachBru2016}
V.~Bach and J.-B. Bru.
\newblock Diagonalizing quadratic bosonic operators by non-autonomous flow
  equation.
\newblock {\em Memoirs of the AMS}, 240(1138):1--122, Mar 2016.

\bibitem{BachPedraMerkliSigal2014}
V.~Bach, W.~de~Siquiera~Pedra, M.~Merkli, and I.~M. Sigal.
\newblock Suppression of decoherence by periodic forcing.
\newblock {\em J. Stat. Phys}, 155(6):1271--1298, Jun 2014.

\bibitem{EngelNagel}
K.-J. Engel and R.~Nagel.
\newblock {\em A Short Course on Operator Semigroups}.
\newblock Universitext. Springer-Verlag, 2006.

\bibitem{Facchi}
P.~Facchi, S.~Tasaki, S.~Pascazio, H.~Nakazato, A.~Tokuse, and D.~A. Lidar.
\newblock Control of decoherence: Analysis and comparison of three different
  strategies.
\newblock {\em Phys. Rev. A}, 71(2):022302, Feb 2005.

\end{thebibliography}
\inputencoding{utf8}

\end{document}